\documentclass[prd, twocolumn, nofootinbib]{revtex4}
\usepackage{graphicx}
\usepackage{dcolumn}
\usepackage{epsfig} 
\usepackage{amsmath}
\usepackage{amsfonts}
\usepackage{graphicx}
\usepackage{amssymb}
\usepackage{color}

\begin{document}
\title{An improved model of redshift-space distortions around voids: application to quintessence dark energy}

\author{Ixandra Achitouv$^{1,2}$\footnote{E-mail:iachitouv@swin.edu.au}}

\address{ 
$^{1}$Centre for Astrophysics \& Supercomputing, Swinburne University of Technology, P.O. Box 218, Hawthorn, VIC 3122, Australia \\
$^{2}$ARC Centre of Excellence for All-sky Astrophysics (CAASTRO), 44 Rosehill St, Redfern, NSW 2016, Australia
}

\begin{abstract}
Using cosmic voids to probe the growth rate of cosmic structure, and hence the nature of dark energy, is particularly interesting in the context of modified gravity theories that rely on the `screening mechanism'. In this work we improve the modelling of redshift-space distortions around voids in the dark matter density field, and thus reduce systematic errors in the derivation of  cosmological parameters. We also show how specific types of voids can be used to better probe the growth rate, using a flexible void finder. We apply our results to test for a `quintessence' type of dark energy vs. a $\Lambda$CDM model, and find a good agreement with the fiducial cosmology after implementing an analytical correction to the radial velocity profiles around voids. We additionally outline characteristic imprints of dark energy in the dark matter velocity distributions around voids. 
\end{abstract} 
\maketitle

\section{Introduction}
Several cosmological observations have provided essential insights into the formation of the cosmic web within the $\Lambda$CDM concordance model (e.g. \citep{Planck2015,SN1a,Cluster,BAO,Lens}). However, a physical interpretation of dark energy is still missing, prompting investigation into alternative cosmological models  (e.g. `dynamical dark energy' \citep{RPCDM,coupledDE}), as well as modified gravity theories \citep{HuSaw,chame,Khoury2004}. Because these alternative interpretations of dark energy are required to have a similar expansion history of the universe to what has been observed, it is often difficult to distinguish them from a simple cosmological constant. In this context, cosmic voids could provide complementary constraints on dark energy. 
\medskip

Indeed, over the last decade, efforts have been made to use cosmic voids in order to challenge the $\Lambda$CDM model (e.g.\citep{Granettetal2009,Lavaux2012,ABPW,Zivick2015}), by looking at Alcock-Paczynski tests \citep{Sutter_APtest,Mao2017}, the integrated Sachs-Wolfe effect (e.g.\citep{Granettetal2009,Kovacs2017}), lensing around voids (e.g.\citep{Melchior2014,Clampitt2014}), and measuring their abundance, their density profiles (e.g.\ \cite{ANP,ABPW,Clampitt2013,Zivick2015}), or looking at the clustering of matter in underdense environments \citep{AB_BAO,Kirauta16}. 

\medskip
Redshift space distortions (RSD) are sourced by the peculiar velocities of galaxies, and are sensitive to the linear growth rate, $f\equiv \frac{d\ln{\delta_m(a)}}{d\ln{a}}$, with $\delta_m(a)$ the growing mode of matter density fluctuations and $a$ the scale factor. On large scales the linear growth rate can be measured by probing the coherent infall or outflow of galaxies, sourced by the gravitational potential, while at smaller scales we can observe the elongation of the clustering signal of galaxies along the line-of-sight (ie: finger of god). The linear growth rate is a powerful cosmological probe, sensitive to both cosmic expansion and to the  gravitational interactions that shape the formation of cosmic structures. It has been inferred using galaxy clustering in redshift space, from many datasets, including the {6 degree Field Galaxy Survey} 6dFGS \citep{Beutler6dF,AB_voids2016}, the 2 degree Field Galaxy Redshift Survey (2dFGRS) \citep{Peacock2001,Hawkins2003, Cole2005}, the Sloan Digital Sky Survey (SDSS) \citep{Tegmark2006}, the WiggleZ Dark Energy Survey \cite{Blake2011}, the Baryon Oscillation Spectroscopic Survey (BOSS) \citep{Reid2012} and the VIMOS Public Extragalactic Redshift Survey (VIPERS) \cite{delaTorre2013}. 
\medskip

However, it was only recently that the growth rate was probed via RSD around voids, in galaxy surveys at low and high redshift (BOSS-CMASS \cite{HamausSDSS} sample, 6dFGS \cite{AB_voids2016} and VIPERS \cite{AH_Viper2016}). In fact, probing the growth rate in underdense regions is particularly interesting when we consider modified gravity models (e.g. $f(R)$ models \citep{HuSaw,chame}), which rely on `screening mechanisms'. Such mechanisms allow for departures from general relativity in underdense regions, meaning that cosmic voids could play a central role to reveal the signature of modified gravity theories \citep{ABPW,Zivick2015,Li2012,Cai2015,IA2016}. Similarly, quintessence dark energy or coupling dark energy models can impact not only the cosmic expansion but also the formation of cosmic structures \citep{Vernizzi2015,DuttaMaor2007,2016Paillas,Adermann2017}  which includes cosmic voids. In addition, a comparison of the growth rate measurement within under-dense regions and within the average density of the universe (\cite{AB_voids2016}) can reveal non-standard cosmological models.
\medskip

 The theoretical framework used to model RSD around voids (e.g. \cite{HamausSDSS,AB_voids2016,AH_Viper2016}) relies on the \textit{Gaussian Streaming Model} (e.g.\citep{Fisher95,Reid2011,KoppUA}) with a linear approximation between the radial peculiar velocity and the underlying density contrast (\cite{Peebles}). However, this simple model has been shown to lead to systematic errors in the inferred growth rate (e.g. \citep{IA2016,HamausSDSS2,Cai2016}). This is why, in this work, we study how to improve the modelling of RSD around voids in order to reduce systematic errors in the inferred cosmological parameters. In particular, by introducing a semi-empirical correction to the mean velocity profile around voids, which depends on the local density contrast, and by using a velocity dispersion that is not constant, we show how we can alleviate systematic errors in the derivation of the growth rate. 
\medskip

We additionally study how the type of void (particularly the density contrast at the ridge of the void) impacts the derived cosmological parameters (e.g. \citep{Cai2016,IA2016,Laresetal17}). Bearing this in mind, particular attention has to be given to the void finder, and the effect of the bias, when identifying voids with different tracers (e.g. \cite{Pollina2017}). To differentiate bias-induced systematic errors from others, when probing RSD in galaxy surveys, we consider in this work the derivation of the growth rate using dark matter mock catalogues. 

\medskip

This paper is organized as follows: in Sec.~\ref{secmocks}, we describe how we build our mock catalogues and identify cosmic voids. In Sec.~\ref{Sec2}, we propose a correction to the radial velocity profiles around voids, as a function of the local density contrast, and we test this correction within our mocks. In Sec.~\ref{sec4}, we apply this correction to infer the growth rate from the clustering around voids in the redshift space mocks. Finally, in Sec.~\ref{DEsec}, we test our framework on a quintessence dark energy model and highlight its imprints in the peculiar velocity distribution within voids. In Sec.~\ref{conclu} we present our conclusions.

\section{Void finder \& Dark Matter mocks}\label{secmocks}
In what follows we employ the DEUS N-body simulations, which have been used for several purposes (e.g.\ \citep{Alimi2010,Courtin2011,Rasera2010,ARSC,AWWR}) and were realized using the RAMSES code \cite{Teyssier2002} for a $\Lambda$CDM model with $(\Omega_m,\sigma_8,n_s,h)=(0.26,0.79,0.96,0.7)$ and a Ratra-Peebles quintessence dark energy model (RPCDM), i.e., $(\Omega_m,\sigma_8,n_s,h)=(0.23,0.66,0.96,0.7)$, where the dark energy equation of state is given by $w(a)=-0.87+0.08(1-a)$. The simulation setup is such that the box length is $5250 \rm Mpc.h^{-1}$ and $5184 \rm Mpc.h^{-1}$ for the $\Lambda$CDM, and RPCDM, respectively, both having a total a number of particles equal to $2048^3$. From each simulation we built 20 dark matter mocks of length $656.25 \rm Mpc.h^{-1}$, randomly sub-sampling the dark matter particles until the density in each mock equals $\bar{n}=0.001 \rm Mpc^{-3}.h^{3}$. We also highlight that these simulations have initially the same density fluctuation normalization, and are consistent with cosmological constraints from WMAP-5yrs (realistic models). Any differences measured in the statistical properties of the dark matter field at $z=0$ are due to the dynamical aspect of the quintessence type of dark energy, which changes the background expansion of the universe but also leaves imprints in the non-linear regime, where cosmic structures form \citep{Alimi2010,Courtin2011,Rasera2010}. 

\medskip

We identify voids using the algorithm presented in \cite{IA2016} (their sec. 2). This void finder has two main advantages. First, unlike void finders based on the watershed concept (e.g.\ \cite{N08,VIDE}), our resulting void density profiles are not too sensitive to the number density of the tracers. This is essential to calibrate the real space (dark matter - void) correlation function, since this quantity should only be sensitive to the fiducial cosmology. This robustness to the number density is due to the algorithm, which seeks for conditions on the density contrast, in a limited number of bins $R_i$. Second, this void finder allows for flexibility in the type of voids that we want to select. This is an important point, as it was discussed in \cite{IA2016}, since the amplitude of the density contrast at the ridges can lead to a systematic error in the growth rate inference when using RSD clustering. 

\medskip
In what follows we will consider 3 different samples of voids ($S_1, S_2, S_3$) that have a different density perturbation at the ridge, $\delta(R=R_v)$, where $R_v$ is the void radius. The voids are defined to satisfy the following conditions:
$\xi_{\rm v-DM}(R_v)>\delta_{\rm thresh}, \xi_{\rm v-DM}(R_v+dR)>\xi_{\rm v-DM}(R_v), \xi_{\rm v-DM}(R_v+dR)<\delta_{\rm max}, \xi_{\rm v-DM}(R<R_{\rm core})<\delta_{\rm core} $. The steps of the algorithm to obtain these conditions are described in sec. 2 of \cite{IA2016}, and the values of the parameters for the 3 different samples of voids are listed in Tab.\ref{Tab1}. In the case of overlapping voids, we only select the largest one.

\begin{tiny}
\begin{table*}[t]
\centering
\begin{tabular}{l c c c c c}
 Void Samples& $\delta_{\rm thresh} $ & $\delta_{\rm max}$ & $\delta_{\rm core}$ & $R_{\rm core} [Mpc.h^{-1}]$  \\ 
\hline \hline
$ S_1$& $ 0$ & $0.3$ & $-0.3$ & $9$  \\
$ S_2$& $ -0.1$ & $0.2$ & $-0.3$ & $9$  \\
$ S_3$& $ 0.15$ & $0.4$ & $-0.3$ & $9$  \\
\hline
\end{tabular}
\caption{Density criteria that defined our 3 different void samples.}
\label{Tab1}
\end{table*}
\end{tiny}

\medskip
In each of the mocks we found voids with a range of radii between $R_v\sim [10-80] \rm Mpc.h^{-1}$. Given the mean density in each of our mocks, we focus our study on voids with a fiducial radius of $R_v=40 \rm Mpc.h^{-1}$, finding a total of $400,512,160$ voids of $40 \rm Mpc.h^{-1}$ in all mocks, for samples $S_1,S_2$ and $S_3$, respectively. 

\medskip

\begin{figure}[ht]
\centering
\includegraphics[scale=0.4]{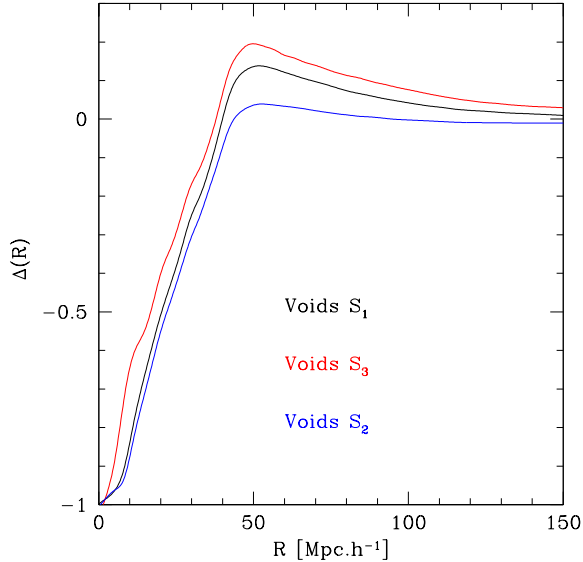} 
\caption{Cumulative density profiles measured in the real space mocks, for the 3 different void samples.}
\label{Fig1}
\end{figure}

Once we have our void catalogues for each sample, and for our two cosmologies, we employ the Landy-Szalay estimator (\cite{LSestimate}) to compute the galaxy-void correlation function, or in our case, the dark matter-void correlation function
\begin{equation}
\begin{split}
\xi_{\rm v-DM}(R)&=\frac{D_{\rm v}D_{\rm g}}{R_{\rm v}R_{\rm g}}\frac{N_{\rm Ran}N_{\rm Ran2}}{N_{\rm DM} N_{\rm Void}} -\\
&\frac{D_{\rm v}R_{\rm v}}{R_{\rm v} R_{\rm g}}\frac{N_{\rm Ran}}{N_{\rm Void}}-\frac{D_{\rm g}R_{\rm g}}{R_{\rm v}R_{\rm g}}\frac{N_{\rm Ran2}}{N_{\rm Void}}+1, 
\end{split}
\end{equation}
where $N_{\rm Void}$ corresponds to the number of voids identified with the required density criteria, and $N_{\rm Ran2}$ corresponds to the number of random set positions that overlap with the voids. The number of pairs separated by a distance $R$ is labelled by $\rm D_g, D_v$ for the dark matter particles and void data, respectively, while  $\rm R_g, R_v$ correspond to dark matter particles and void pairs computed from the random distributions. In Fig.\ref{Fig1} we show the cumulative density profiles (i.e. Eq.\ref{EqDelta}) for the 3 different void samples, by stacking void profiles in all the real space mocks. As expected, by increasing $\delta_{\rm thresh}$, the amplitude at the ridge follows suit. Note that we implicitly assume spherical symmetry to compute the 1-dimensional density profiles.

\medskip
Finally, we use the same dark matter density mocks to build the redshift space (RS) mock catalogues, and we repeat our void identification process for our 3 samples of voids. To do so we shift the real space dark matter particle positions, $\mathbf{r}$, to $\mathbf{s}$, using the flat sky approximation:

\begin{equation}
\mathbf{s}=\mathbf{r}+\frac{v_p(\mathbf{r})}{H_0}\mathbf{u}_r,
\end{equation}
where $\mathbf{u}_r$ is the unitary vector along the line the sight (here the $z$-coordinate), and $v_p\equiv\mathbf{v}.\mathbf{u}_r$ is the peculiar velocity of the particles along the $z$-direction.

\section{Accurate modelling of RSD around voids}\label{Sec2}
\subsection*{Theory}
In the Gaussian streaming models (GSM) (e.g.\citep{Fisher95,Reid2011,KoppUA}) the void-matter correlation function in the local universe can be expressed as a convolution between the real space correlation function and the probability density distribution of the velocities of dark matter particles/galaxies around voids:
 \begin{equation}
 \begin{split}
 \xi_{vg}(\sigma,\pi)&=\int (1+\xi_{\rm v-DM}^{1D}(y)) \times\\
 & P\left(v-v_g(y) \left[\pi-\frac{v/H_0}{y}\right]\right)dv -1,
\label{GSMeq}
\end{split}
 \end{equation}
where $v_g(r)\equiv <\delta_v v>$ is the mean radial velocity profile of dark matter around  voids $\delta_v$, $\pi$ is the line of sight co-ordinate, $y=\sqrt{\sigma^2+(\pi-v/H_0)^2)}$, with $\sigma$ the perpendicular direction to the line of sight, and $P(v) dv$ is the stochastic velocity distribution of matter within a group (reproducing the small scale elongation along the line of sight). The velocity distribution can be approximate by a Gaussian\footnote{In \cite{AB_voids2016} the authors have shown that an exponential distribution would lead to similar constraints of the growth rate}: 

\begin{equation}
P(v)dv=\frac{1}{\sqrt{2\pi\sigma_v}} \exp\left[-\frac{v^2}{2\sigma_v}\right] dv,\label{PDFv}
\end{equation}
where $\sigma_v(R)$ is the velocity dispersion parameter. Previous works that have probed the growth rate with RSD around voids in galaxy surveys (e.g. \citep{HamausSDSS, AB_voids2016, AH_Viper2016}) have used, for the radial mean velocity profile, the linear approximation (\cite{Peebles}) given by $v_{g}(R)=-1/3 H_0 R\Delta(R) f$, while $\sigma_v$ is  described as a constant parameter. The cumulative density contrast for spherical symmetry is defined as

\begin{equation}
\Delta(R)=\frac{\int \mathbf{dV} \delta_g}{4/3 \pi R^3}=\frac{3}{R^3}\int_{0}^{R} dr \; r^2 \delta_{g}(r),\label{CDP}
\end{equation}
where $\delta_g\sim b\delta_{\rm DM}$ is the galaxy density contrast at a given scale, and is approximately equal to the linear bias $b$ times the dark matter density contrast $\delta_{\rm DM}$. In what follows we use the notation $\xi_{vg}$ for the 2D voids-dark matter particles correlation function.
\medskip

In this work we will show that the linear approximation is not good enough to accurately describe the radial velocity of dark matter particles (hence galaxies) around voids, $v_g$, especially when these voids have a high amplitude at the ridge. More importantly, we will discuss how this linear approximation, and how choosing a constant velocity dispersion, can lead to systematic errors in the derivation of the growth rate. Hence, in what follows, we will improve the modelling of the radial velocities of dark matter particles around voids based on a semi-empirical formula that we introduce. 

\medskip
We start by reviewing the well known Zeld'dovitch approximation to derive the mean velocity profile around voids. Using the standard notation (e.g. Eq.98 in \cite{Bernardeau2001}), we have

\begin{equation}
\textbf{x}(\textbf{q})=\textbf{q}-D_1\nabla_q\phi^{(1)}+D_2\nabla_q\phi^{(2)},  
\end{equation}
with $\textbf{x}$ the Eulerian position of a galaxy (or the dark matter particle), 
$\textbf{q}$ the initial position, $\phi^{(1)},\phi^{(2)}$ the Lagrangian potentials up to second-order, $D_1$ the usual growth factor, and $D_2$ the second order growth factor, which for $0.01\leq \Omega_m \leq 1$ can be approximated by $D_2 \simeq -3/7 D_{1}^{2}\Omega^{-1/143}$ to better than $0.6\%$ (\cite{Bouchet95}) and in what follows we take $\Omega^{-1/143} \simeq 1$. For the velocity field (e.g. Eq.99 in \cite{Bernardeau2001}), 

\begin{equation}
\mathbf{v_q}=-D_1 f_1 a H \mathbf{\nabla}_q \phi^{(1)}+D_2 f_2 a H \mathbf{\nabla}_q \phi^{(2)}, \label{v2LPT}
\end{equation}
where $f_1\equiv f$ is the standard linear growth rate that can be approximate for a flat universe with a cosmological constant by $f_1\simeq \Omega_{m}^{5/9}$, $f2 \equiv d\ln D_2/d\ln a$ can be approximated by $f_2\simeq 2 \Omega_{m}^{6/11}$ \cite{Bouchet95}, and $a H$ is the conformal expansion rate. The Poisson equations of the potentials also give $\nabla_{q}^{2} \phi^{(1)}=\delta^{(1)}$ and $\nabla_{q}^{2} \phi^{(2)}=-1/2\sum_{i,j}[(\nabla_{i,j}\phi^{(1)})^2-(\nabla^2\phi^{(1)})^2]$ (e.g. Eq. 29-30 of \cite{Chan14} or Eq.103 of \cite{Bernardeau2001}). 
 
\subsubsection*{The linear regime} 
 
The linear regime corresponds to the Zel'dovitch approximation, $\delta(\mathbf{x,t})=D(t)\delta(\mathbf{x})$, and the Lagrangian time derivative of the peculiar velocity is equivalent to the Eulerian time derivative. Hence, Eq.\ref{v2LPT} becomes in Eulerian space, at first order (1PT) and for $z=0$, (see e.g. \cite{Peebles},\cite{Bernardeau2001}):
 
 \begin{equation}
 \mathbf{v^{(1)}}=- f_1 H_0 \mathbf{\nabla} \phi^{(1)}, \label{Eq1PT}
 \end{equation}
with $\nabla^2 \phi^{(1)}=\delta^{(1)}$. Taking the divergence of Eq.\ref{Eq1PT} and using spherical coordinates, we deduce the radial peculiar velocity of galaxies, or dark matter particles, to be:

\begin{equation}
\frac{1}{r^2}\frac{\partial r^2 v(r)}{\partial r}=-f _1H_0 \delta^{(1)},
\end{equation} 
 which can be integrated to
 
 \begin{equation}
v^{(1)}(R)=-f_1 H_0 \frac{1}{R^2} \int_{0}^{R} r^2\delta^{(1)}(r) dr.\label{1PT}
\end{equation} 

Hence, the average radial velocity profiles of galaxies/dark matter around voids becomes

 \begin{equation}
\left<\delta_v v^{(1)}\right>\equiv v_g(R)=-f_1 H_0 \frac{1}{R^2} \int_{0}^{R} r^2 \left<\delta^{(1)} \delta_v\right> dr,
\end{equation} 
with $\left<\delta^{(1)} \delta_v\right> \equiv \xi_{\rm DM-v}$. Using the cumulative density contrast, we recover $v_{g}(R)=-1/3 H_0 R\Delta(R) f$, where $\xi_{\rm DM-v}$ replaces $\delta_g$ in Eq. \ref{CDP}, and $v_g(r)$ is the radial velocity profile of dark matter particles around voids. 

\subsubsection*{Semi-empirical correction of the radial velocity profile}

The second order correction in Eulerian space, starting from Eq.\ref{v2LPT}, involves more computation (see for instance \citep{Peebles,Hui1996,Catelanetal1995}). Essentially, the divergence of the second order Eulerian peculiar velocity is a sum of three terms: one proportional to the original perturbation $[D_1\delta^{(1)}(q)]^2$, one non-local term that involves the initial perturbation $(\mathbf{\nabla} \phi^{(1)}.\mathcal{\nabla})\delta^{(1)}$, and one is the velocity shear contribution. In particular see Eq. 43 of \cite{Catelanetal1995} for a derivation of the 3 different terms.

\medskip
 However, we stress that this derivation is for a density perturbation $\delta_g$ at a given point in space. As we want to apply this derivation to galaxies or dark matter around voids, we need to take into account the average $v_{g}(r)=\left<\delta_v v\right>$. In principle, it is possible to implement such corrections to peculiar velocities of galaxies around voids, but it would be challenging, especially the velocity shear contribution. In addition, the resulting corrections would be rather unwieldy. 

\medskip 
This is why instead we propose a semi-empirical expression for the velocity profile around voids:

\begin{equation}
v_g(R)=-f H_0 \frac{R}{3}\Delta(R) +\varepsilon \frac{3}{7} f_2 H_0 \frac{R}{6}\Delta^{(2)},\label{Eq2PTok}
\end{equation}
with 

\begin{eqnarray}
\Delta^{(2)}\equiv \frac{3}{R^3}\int_{0}^{R} r^2 \xi_{vg}^{2} dr,\\
\Delta \equiv \frac{3}{R^3}\int_{0}^{R} r^2 \xi_{vg} dr, \label{EqDelta}
\end{eqnarray}
and $\varepsilon=+1$ when $\Delta>0$, and $\varepsilon=-1$ when $\Delta<0$, in order to improve the linear prediction of the velocity profiles (as we will see shortly). This expression is motivated by the spherical evolution of a perturbation \cite{Bernardeauspherical} where the second order correction with an extra-factor $1/2$ corresponds to one of the terms of the full Eulerian second order perturbation \cite{Catelanetal1995}:
 
\begin{equation}
\nabla.\mathbf{v}^{(2)}=-\frac{3}{7}f_2H_0(\delta^{(1)})^2+ \rm other \; terms.
\end{equation} 
However in our case, we approximate  $\left<\delta_g^2\delta_v\right>\sim \left<\delta_g\delta_v\right>^2$, to use the correlation function $\xi_{vg}$ in our proposed semi-empirical velocity profile around voids.

\begin{figure}[ht]
\centering
\includegraphics[scale=0.43]{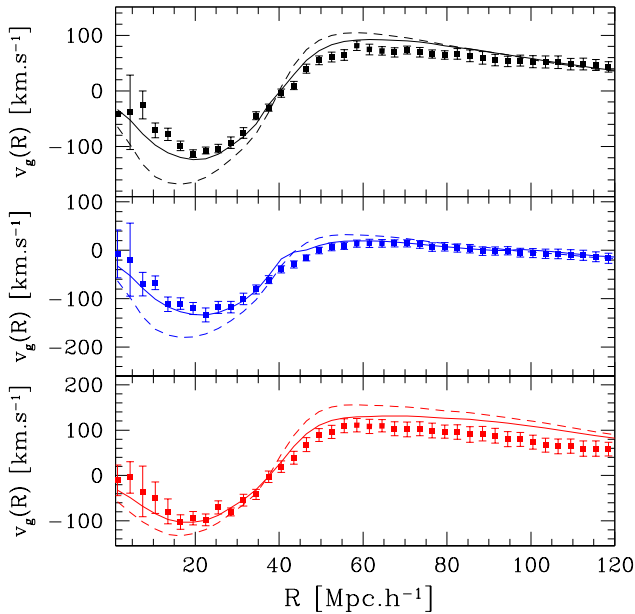} 
\caption{Velocity profiles measured around voids (squares) for the 3 different void samples: $S_1,S_2,S_3$, from top to bottom panels, respectively. The  dashed curves correspond to the linear approximation theory (1PT), while the solid curves correspond to Eq.\ref{Eq2PTok}.}
\label{Fig2}
\end{figure}

\medskip
In Fig.\ref{Fig2}, we compare our prediction for the 3 different void samples. The prediction for $v_g(R)$ using in the linear regime corresponds to the dashed curves, while the solid curves correspond to our proposed semi-empirical correction Eq.\ref{Eq2PTok} and the measured velocity profiles are shown by the squares. This correction avoids large systematic errors in the modelling of the velocity profiles around voids. We also note that the agreement for the void sample $S_3$ is not as good as for the samples $S_1$ and $S_2$. This is directly due to the fact that the assumption of the perturbation theory: $\mid{\Delta^{(2)}}\mid <\mid {\Delta} \mid<< 1$ is more accurate for voids $S_1$ and $S_2$.

\subsection*{Velocity dispersion behaviour}\label{secPDFv}
We now turn to the other assumptions of the Gaussian streaming model. As we have already mentioned, previous cosmological constraints made on the growth rate with RSD around voids \citep{HamausSDSS, AB_voids2016, AH_Viper2016} have used a Gaussian distribution with a constant velocity dispersion, described as a nuisance parameter. As we will outline in sec.\ref{DEsec}, the velocity dispersion in fact carries the imprint of dark energy. 

\medskip

\begin{figure}[ht]
\centering
\includegraphics[scale=0.43]{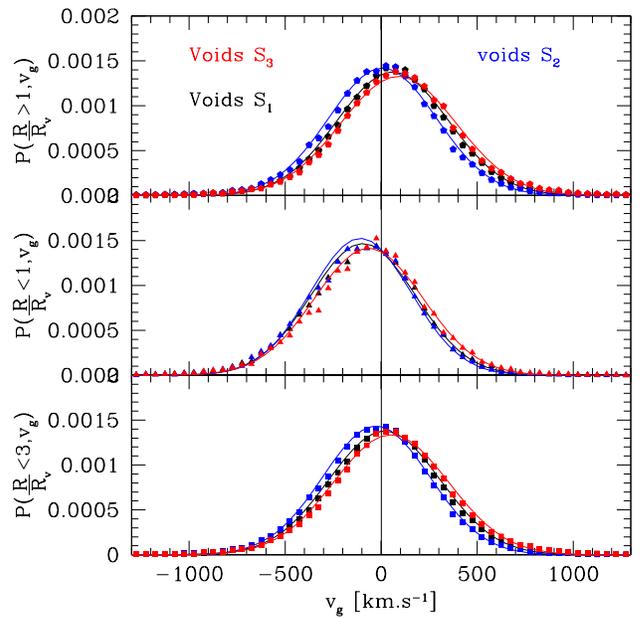} 
\caption{Velocity profile distribution of dark matter particles measured around voids in the real space mocks, for the 3 different void samples: $S_1,S_2,S_3$ (black, blue and red, respectively). The different panels show different radial ranges relative to the void centre: at the inner part of the voids (middle panel triangles), after the ridge of the voids (upper panel diamond) and at all scales (lower panel squares). The  different curves correspond to Gaussian distributions with a mean value fixed by the mean distribution of the velocities at that scale. The standard deviation of the best fitting Gaussian is $\sigma_{\rm eff}=0.9\sigma_v$, where $\sigma_v$ is the standard deviation measured from the distribution at that scale.}
\label{FigPDFvp1}
\end{figure}

In Fig.\ref{FigPDFvp1} we show the probability density function (PDF) of the dark matter particles around voids, for our different voids samples $S_1,S_2,S_3$ (black, blue and red, respectively) and at different radii (diamonds, triangles and squares). In the inner part of the voids ($R/R_v<1$, triangles, middle panel), the mean value of the PDF is negative since matter is being pushed towards the ridge of the voids. Beyond the ridge ($R/R_v>1$, diamonds, top panel) the average of the distribution is positive since voids $S_1$ have a ridge, causing a positive bump in the velocity profiles (i.e. $v_g(R)\propto \Delta(R)$). If we consider all scales ($R/R_v<3$, squares, lower panel), then we recover a mean value which is approximately zero: $\Delta(R>>R_v)= 0)$. For each distribution we measure the mean and the standard deviation $\sigma_{v}$. The solid curves show Gaussian distributions with a mean sets by the average of the PDF measurement, and a standard deviation $\sigma_{\rm eff}=0.9\sigma_v$. This effective standard deviation is because the measured PDFs  are leptokurtic (excess kurtosis), while the skewness is negligible. This is a direct consequence of using several voids to build these PDFs, each void contributing $v_g$ values that are Gaussian-distributed, with standard deviation $\sigma_{v}(R)$. Hence, the combination of these Gaussian PDFs leads to an excess of kurtosis. Interestingly the effective standard deviation is insensitive to the type of voids we choose, and is quasi independent of the distance from the void center. 

\medskip
\begin{figure}[ht]
\centering
\includegraphics[scale=0.43]{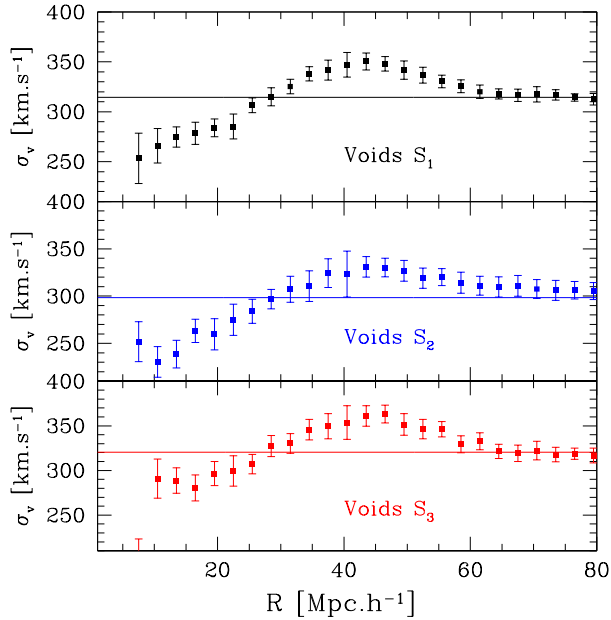} 
\caption{Velocity dispersion profiles of dark matter particles measured around voids in the real space mocks, for the 3 different void samples. The solid lines correspond to the mean velocity dispersion measured over the range $R=[5,80]\rm Mpc.h^{-1}$. }\label{Figsigrealsp}
\end{figure}

In Fig.\ref{Figsigrealsp} we show the measured velocity dispersion around voids, $\sigma_v(R)$, from the real space mocks, for voids sample $S_1,S_2,S_3$ (black, blue and red respectively). The solid lines correspond to the mean value of the velocity dispersion over the range $R=[5,80] \rm Mpc.h^{-1}$. It is obvious that the velocity dispersion is not a constant with radius, as was already observed (e.g. \cite{Hammaus2015JCAP}), and so one might wonder how modelling it as a constant in the Gaussian steaming model affects the derivation of the growth rate. 

\subsection*{Impact on the 2D correlation function and velocity profiles}
In this section we test the (dark matter-void) correlation function in redshift space, as well as the velocity profiles, for 2 cases: case 1, using the linear approximation,  and case 2, using our proposed formula (Eq.\ref{Eq2PTok}), both with the measured velocity dispersion scaled as $\sigma_v\rightarrow 0.9\sigma_v(R)$ and for the fiducial cosmology $f\sigma_8\approx 0.376$, with void sample $S_1$.

\medskip
To model the velocity profile in redshift space (along the line of sight $\pi$, and perpendicular to the line of sight $\sigma$) we use:
\begin{equation}
v_{g}^{RS}(\sigma,\pi)=\int v_g(y) \times P\left(v-v_p(y) \left[\pi-\frac{v/H_0}{y}\right]\right)dv,\label{Eqvp2PT}
\end{equation}
where again, $v_g(R)$ is the averaged velocity profile of dark matter predicted by Eq.\ref{Eq2PTok}, $y=\sqrt{\sigma^2+(\pi-v/H_0)^2)}$, and $P(v) dv$ is the same velocity distribution of Eq.\ref{GSMeq}.

\begin{figure}
\centering
\begin{tabular}{cc}
\includegraphics[scale=0.4]{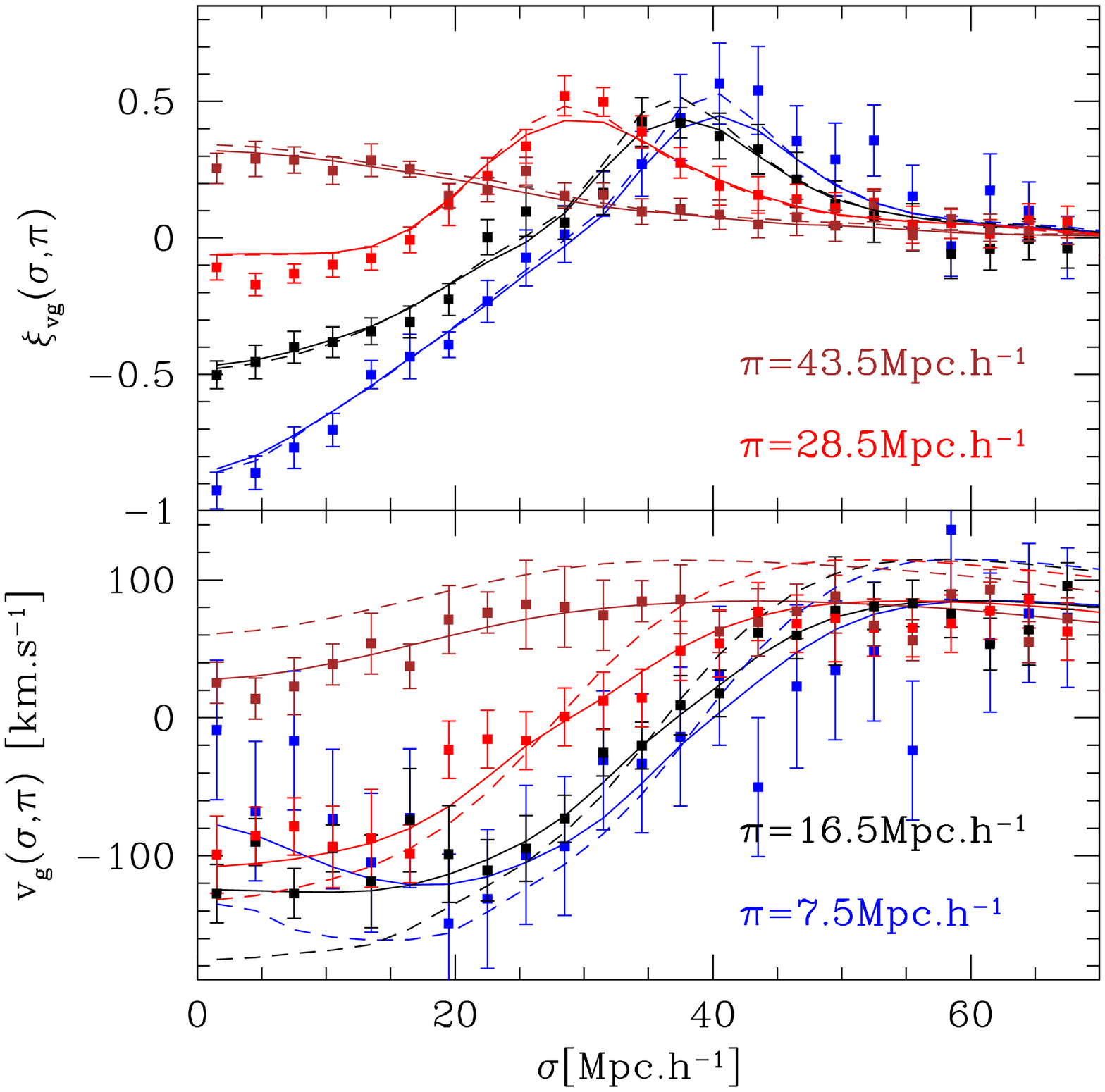}\\
\includegraphics[scale=0.4]{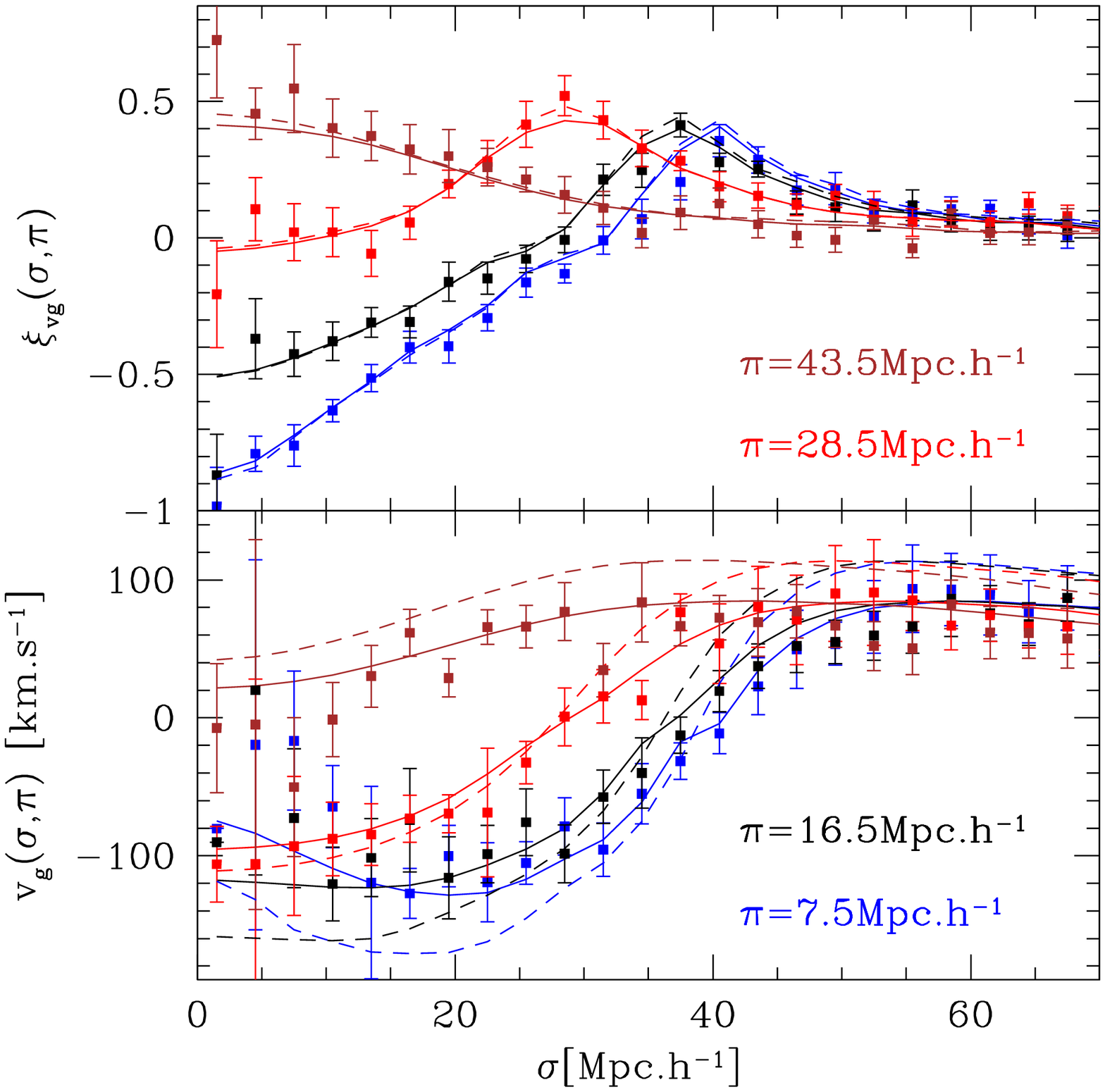}
\end{tabular}
\caption{2D (void-dark matter) correlation function and 2D velocity profiles around voids, for sample $S_1$ (top plot shows fixed value along the line-of-sight, bottom plot shows fixed values across the line-of-sight). In each plot, the squares are the measured values in the mocks, the solid curves correspond to the fiducial cosmology prediction with Eq.\ref{Eq2PTok} plugged into Eqs.\ref{GSMeq}, \ref{Eqvp2PT}, while the dashed curves correspond to the linear approximation of the velocity profile.}
\label{Fig2Dcomp}
\end{figure}
\medskip
In Fig.\ref{Fig2Dcomp} we perform a comparison between our two cases (case 1 - dashed curves \& case 2 - solid curves) and the mocks, at fixed bins of $\pi$ (top plot) and fixed bins of $\sigma$ (bottom plot). We observed that the velocity profiles in redshift space show a clear preference for Eq.\ref{Eq2PTok} compared to the linear approximation. On the other hand, it is not possible to see a clear preference of the model by looking at the correlation function. This is why in the next section we will perform a \textit{Markov Chain Monte Carlo} to analyse the correlation functions.

\section{Reducing systematics errors of the growth rate measurement}\label{sec4}
To test the derivation of the growth rate from our void samples, we perform a Metropolis-Hastings \textit{Markov Chain Monte Carlo} (MCMC) analysis for the parameters $\Theta=(\sigma_v,f\sigma_8)$. The  likelihood of each void sample is computed from
\begin{equation}
\chi^2(\Theta) = \sum_{\sigma,\pi} \left(\frac{\xi^{\rm
    \rm mocks}(\sigma,\pi)-\xi^{\rm theo}(\Theta,\sigma,\pi)}{\sigma_{\rm mocks}
  (\sigma,\pi)}\right)^2.
\label{1Dchisq}
\end{equation}

We also compute the standard deviation across the 20 mocks, $\sigma_{\rm mocks}(\sigma,\pi)$, assuming no correlation between the bins. To model our measurement we use Eq.\ref{GSMeq}, where $\xi_{\rm v-DM}^{1D}$ is calibrated using the real space mocks mean measurement. The $\chi^2$ is performed from ranges between $R_{\rm cut} - \rm 75 Mpc.h^{-1}$ in $(\sigma,\pi)$, with bins of width $3 \rm Mpc.h^{-1}$. Note that for each step, we generate a value of $(\sigma_v, f\sigma_8)$, not $f_2$. To get the corresponding $f_2$ value, we use the approximation of  \cite{Bouchet95}, i.e., $f_2\simeq 2 (f^{9/5})^{6/11}$. In principle this approximation is valid for a non-zero cosmological constant but in what follows we will also use it for our realistic model of quintessence dark energy. This will be justified in sec.\ref{DEsec} by ensuring that the predicted velocity profile also reproduces the measured one, for the RPCDM model.


\subsection*{The linear approximation with constant velocity dispersion}

We start by investigating the derivation of the growth rate using the standard linear approximation for the mean velocity profile (1PT), Eq.\ref{Eq2PTok}, with $\Delta^{(2)}=0$, and a constant velocity dispersion, as it has been previously used in \citep{HamausSDSS, AB_voids2016, AH_Viper2016}.

\begin{figure}[ht]
\centering
\includegraphics[scale=0.45]{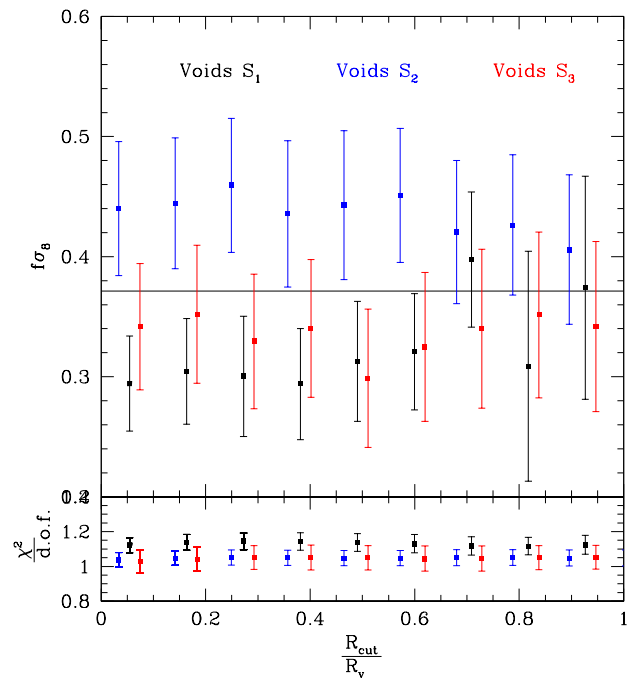}
\caption{Best fitting values of $f\sigma_8$ as function of the fitting range $R_{\rm cut}/R_v$, for the 3 different void samples, using the linear approximation for the velocity profile and a constant velocity dispersion. Different void samples are not consistent with one another at 1-sigma for $R_{\rm cut}/R_v<0.7$, in the inferred values of $f\sigma_8$.}\label{BF_H1}
\end{figure}

\medskip
In Fig. \ref{BF_H1} we show the inferred values of $f \sigma_8$ for our 3 different void samples. The mean value of the best fitting parameters over the mocks corresponds to the best fitting value of the mock mean, and we use the scatter of the best fitting values across the mocks to estimate the uncertainties on the growth rate. The $x$-axis shows the cut in the fitting range $R_{\rm cut}=\sqrt{\sigma^2+\pi^2}$, when performing the $\chi^2$. This is motivated by \cite{Cai2016}, who show that linear theory may break down at the center of the voids, where $\delta_v(R\rightarrow0)\sim -1$. The solid line corresponds to the fiducial cosmology $f\sigma_8\approx 0.376$, and the lower panel shows the average of the reduced $\chi^2$ over the mocks. 
\medskip

\begin{figure}
\centering
\includegraphics[scale=0.45]{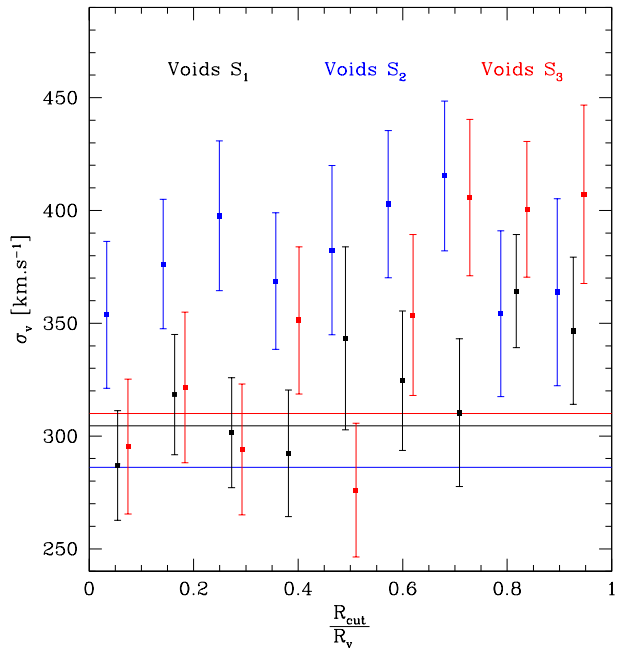}
\caption{Best fitting values of $\sigma_v$ as function of the fitting range $R_{\rm cut}/R_v$, for the 3 different voids samples, using the linear approximation for the velocity profile and a constant velocity dispersion. The solid lines show the measured mean value of the velocity dispersion from the redshift space mocks, over the fitting ranges $[0,75]\rm Mpc.h^{-1}$ where we performed the MCMC analysis}\label{BFsig}
\end{figure}

Overall we see that the different void samples show systematic errors in the growth rate derivation for $R_{\rm cut}/R_v<0.7$. This is not surprising, for two reasons. First, we have seen that the linear approximation for the velocity profile around voids poorly reproduces the measurements within the inner part of the voids. Second, using a constant velocity dispersion is itself wrong for $R/R_v<1.5$. In Fig.\ref{BFsig} we show the best fitting values for the velocity dispersion for the 3 different void samples, as a function of the fitting range $R=[R_{\rm cut}-75]\rm Mpc.h^{-1}$. 

The solid lines show the measured mean value of the velocity dispersion, measured in the real space mocks, over the fitting range $[0,75]\rm Mpc.h^{-1}$. As we knew from Fig.\ref{Figsigrealsp}, as $R_{\rm cut}$ increases, we expect the best fitting values of $\sigma_v$ to increase as well. This is roughly what we observe for voids $S_1,S_3$. For $R_{\rm cut}/R_v=0$ we should recover the measured mean values of the velocity dispersion. This is indeed the case for voids $S_1,S_3$, but not for voids $S_2$, which shows a systematically higher value of $\sigma_v$.

\subsection*{Semi-empirical correction to the velocity profiles}
\begin{figure}
\centering
\includegraphics[scale=0.45]{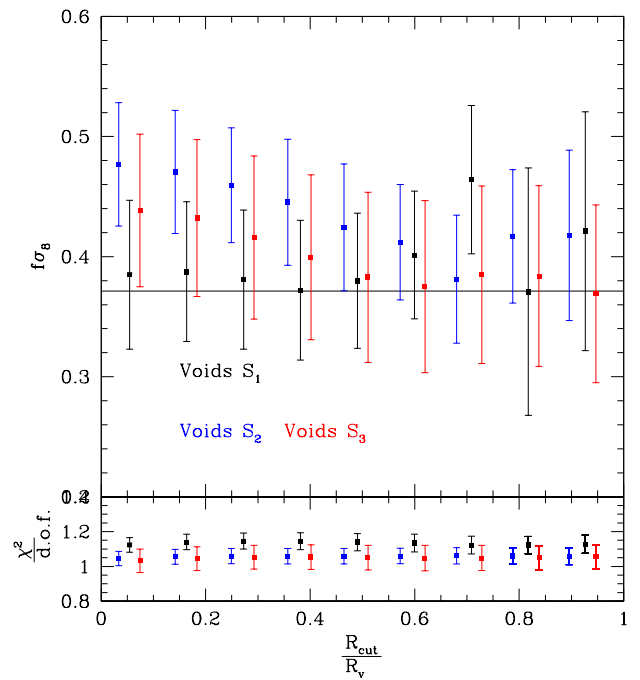}
\caption{Best fitting values of $f\sigma_8$ as function of the fitting range $R_{\rm cut}/R_v$ for the 3 different voids samples, using Eq.\ref{Eq2PTok} for the velocity profile and setting the velocity dispersion to the measurement of the real space mocks: $\sigma_v(R)$. Different void samples are now consistent with one another at 1-sigma in the inferred values of $f\sigma_8$.}\label{BF2PTf}
\end{figure}

We now repeat our analysis, but this time applying our proposed correction to the real space velocity profiles (Eq.\ref{Eq2PTok}). In addition, we now use as part of the model the real space measurement of the velocity dispersion $\sigma_v(R)$ (see Fig.\ref{Figsigrealsp}), and plug it into the Gaussian PDF of Eq.\ref{PDFv}, with the rescaling $\sigma_v\rightarrow 0.9\sigma_v$ to account for the leptokurtic distribution we probed in sec.\ref{secPDFv}. Hence, we only fit for the growth rate factor. 

\medskip
The results of our MCMC analysis can been seen in Fig.\ref{BF2PTf}. We can see that all void samples are now consistent with one another in the inferred values of $f\sigma_8$, and within 1-$\sigma$ agreement with the fiducial cosmology for $R_{\rm cut}/R_v>0.5$. 

\medskip
Finally, in the case of $S_1,S_3$, there is a negligible sensitivity of the inferred growth rate to the fitting range, compared to voids in $S_2$. The sensitivity of the ridge to the inferred cosmology was previously pointed out in \cite{IA2016}, and one might take advantage of selecting voids with a particular shape to avoid systematic errors, without using a cut in the fitting range $R_{\rm cut}$, as it was suggested in \cite{Cai2015}. This is why in the next section we will focus our analysis on the void sample $S_1$. 

\section{Probing quintessence dark energy with RSD around voids}\label{DEsec}

In this section we study the impact of a  Ratra-Peebles`quintessence' type of dark energy \cite{Raserasim} on RSD around voids. This model changes the expansion history of the universe and the formation of cosmic structures \citep{,Courtin2011,Rasera2010,ARSC,AWWR}. Previous works done with this simulation have not investigated the impact on the formation of cosmic voids. The simulations and the model are discussed in \cite{Alimi2010}. The main characteristics of our mocks are summarised in sec.\ref{secmocks}. We consider mocks $S_1$ to test the imprint of dark energy (see Tab \ref{Tab1}). This is motivated by our previous results, showing that mocks $S_1$ have very small systematic errors in the derivation of the growth rate (insensitive to $R_{\rm cut}$). Hence we repeat the identification of voids with characteristic $S_1$ in the RPCDM mocks (real space and redshift space), and then we compute the real/redshift space void-galaxy correlation functions.

\subsection*{Imprint of dark energy on the velocity profiles around voids}

\begin{figure}[ht]
\centering
\includegraphics[scale=0.42]{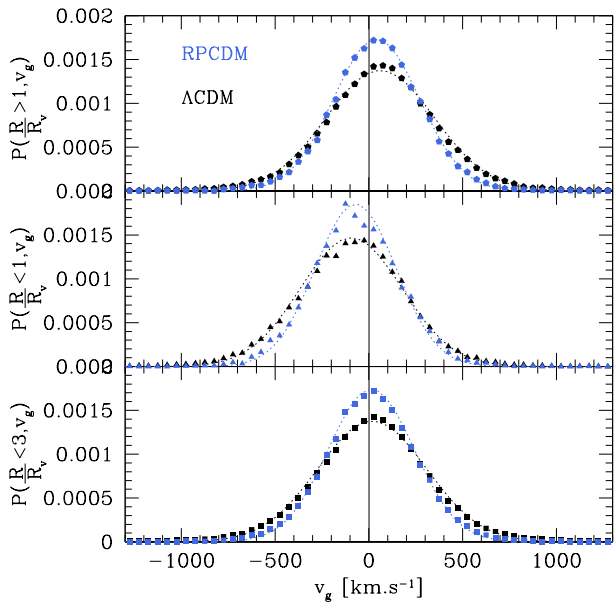} 
\caption{Probability density function of velocities of dark matter particles around voids $S_1$, for $\Lambda$CDM, and RPCDM models. The different panels correspond to the distribution of velocities beyond the ridge of the voids (top panel), within the voids (middle panel), and at all scales (bottom panel).}
\label{FigPDFvp}
\end{figure}

Once again we start by measuring the distribution of the dark matter velocities around voids at different radii from the center. Fig.~\ref{FigPDFvp} shows the comparison between the distribution in the RPCDM mocks (blue distributions) and the $\Lambda$CDM ones (black distributions). The different panels correspond to different scales relative to the center of the voids: at the inner part of the voids (middle panel,  triangles), after the ridge of the voids (upper panel, diamond) and at all scales (lower panel, squares). The  different curves correspond to Gaussian distributions with a mean value fixed by the mean distribution of the velocities at that scale. The standard deviation of the curves is $\sigma_{\rm eff}=0.9\sigma_v$, where $\sigma_v$ is the standard deviation measured from the distribution at that scale. This is again to account for the excess kurtosis discussed in Sec.\ref{secPDFv}. 

\medskip
It is striking to see that the variance of these distributions carries the imprint of dark energy. Indeed, the RPCDM variance ($\propto \sigma_{v}^{2}$) is smaller than that of the $\Lambda$CDM PDFs. This means that the PDFs are more deterministic: there are less non-linear gravitational interactions occurring in the RPCDM model. This is because quintessence dark energy models experience a less efficient deceleration period during the matter dominated era \cite{Alimi2010} ($q=1/2(1+3w)$, with $-1\leq w<0$), such that the mass assembly of matter is overall reduced for a quintessence model of dark energy (and thus is $\sigma_8$).
One may argue that a quintessence dark energy model with a normalisation $\sigma_8$ at $z=0$ matching the $\Lambda$CDM simulation, would lead to similar PDFs of the velocity profiles. However, in this case, the initial density perturbation probed by the Cosmic Microwave Background would rule out this model \cite{Planck2015}, unlike the model we consider here \cite{Alimi2010}. 
 Hence differences in the properties of the cosmic structures at $z=0$, including $\sigma_8$, compared to $\Lambda$CDM are due to dark energy. 

\medskip
Overall these distributions support again the idea that studying $\sigma_v(R)$ could potentially be interesting to probe the growth rate assuming a particular cosmology. 
\medskip

\begin{figure}[ht]
\centering
\includegraphics[scale=0.42]{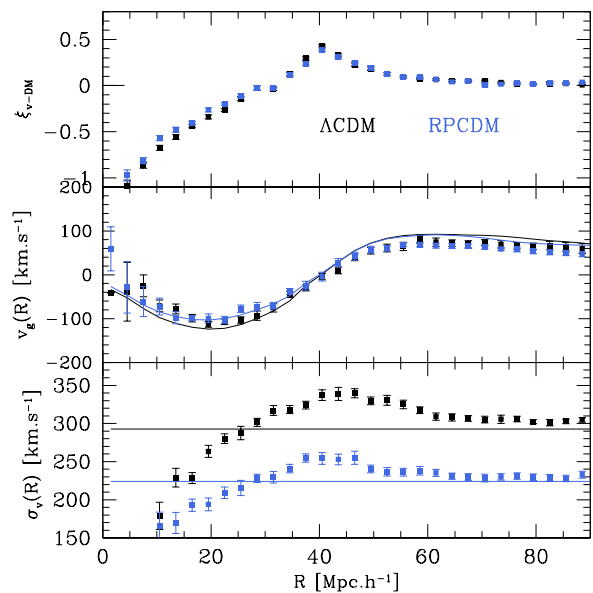} 
\caption{From top to bottom: averaged correlation function measured in real space mocks, mean velocity profiles \&  velocity dispersions.}\label{Figcomp}
\end{figure}

In Fig.\ref{Figcomp} we can see the main characteristics of voids $S_1$ for the $\Lambda$CDM/RPCDM models (black/blue squares and curves).  As expected, the averaged 1D correlation functions $\xi_{v-\rm DM}^{1D}(R)$ (upper panel) measured in the real space mocks, are indistinguishable from one another, as the averaged velocity profiles (middle panel). This is due to our void finder, which selects voids that match the $S_1$ density profile (see sec.\ref{secmocks}). The middle panel shows the measurement of the velocity profiles for the two cosmologies, as well as the prediction of Eq.\ref{Eq2PTok}. The lower panel shows the velocity dispersions. Here we clearly see the imprint of the dark energy model, as we have previously observed in Fig.~\ref{FigPDFvp}. The black/blue lines correspond to the averaged values of the velocity dispersion over the interval $S=[0,90]\rm Mpc.h^{-1}$. Note that our correction in Eq.\ref{Eq2PTok} also matches the trend of the simulation for the quintessence model, using the numerical approximation of \cite{Bouchet95} for $f_2$. 
\medskip

It is quite interesting that the velocity dispersion carries the imprint of dark energy. From an observational point of view, it is quite challenging to obtain direct measurements of peculiar velocities, and a fortiori of the velocity dispersion, although upcoming surveys such as TAIPAN \cite{TaipanWP} might provide a large sample of peculiar velocities at low redshift. Nevertheless, it would be interesting to use RSD to probe both ($\sigma_v,f\sigma_8$) as physical parameters, or fixing $\sigma_v(R)$ by calibrating it to the fiducial cosmology, as we do for $\xi_{v-DM}^{1D}(R)$, which enters into Eq.\ref{GSMeq}.

\subsection*{Probing the growth rate}
We now turn to the derivation of the growth rate for the RPCDM cosmology. We start by repeating the previous analysis we performed for the $\Lambda$CDM voids $S_1$, allowing $(f\sigma_8,\sigma_v)$ as free parameters in the MCMC analysis, with the linear approximation for the velocity profile (1PT). Then we set the velocity dispersion to the measured one in the real space mocks, and use Eq.\ref{Eq2PTok}. The resulting best fitting parameters on the redshift space mocks for $\Lambda$CDM/RPCDM are given in Tab.\ref{Tab4}.

\begin{tiny}
\begin{table}[t]
\begin{tabular}{l c c c c}
 & RPCDM & $\Lambda$CDM\\ 
\hline \hline
1PT $f\sigma_8$ & $0.21\pm 0.03$ & $0.29\pm 0.03$  \\
1PT $\sigma_v$  & $156\pm45$ & $288\pm50$ \\
1PT $\chi^2/d.o.f.$ & 717/625 & 763/625\\
\hline
this work $f\sigma_8$ & $0.27\pm0.03$ & $0.38\pm0.03$ \\
this work $\chi^2/d.o.f.$ & 716/625 & 748/625\\
\end{tabular}
\caption{Parameter constraints on ($f\sigma_8, \sigma_v$) for RPCDM/$\Lambda$CDM, obtained from fitting to the redshift space mocks (dark matter-void correlation function) $\xi_{vg}$ with Eq.\ref{GSMeq}, in the case where we use the linear approximation (1PT). This work corresponds to a velocity profile model by Eq.\ref{Eq2PTok} with the velocity dispersion set to the measured $\sigma_v(R)$. The fiducial cosmology is $f\sigma_8= 0.289,0.376$ and the mean velocity dispersion measured in the real space mocks is $\sigma_v=225,305 \rm km.s^{-1}$ over the fitted range $[0,75]\rm Mpc.h^{-1}$ for the RPCDM/$\Lambda$CDM respectively. }
\label{Tab4}
\end{table}
\end{tiny}

\medskip
As we have previously found, the semi-empirical correction to the velocity profile, combined with a physical description of the velocity dispersion (as opposed to a constant) is essential to recover the fiducial value of the growth rate. This illustrates that RSD around voids can be a good complementary test to probe alternative dark energy models (quintessence in this case). 
\begin{figure}[ht]
\centering
\includegraphics[scale=0.6]{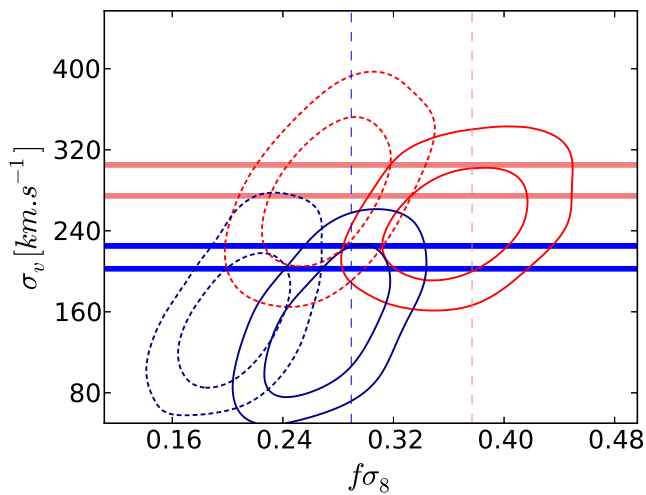} 
\caption{Ellipse constraints on $(f\sigma_8,\sigma_v)$,  derived from the dark matter-void correlation function in our  RPCDM/$\Lambda$CDM redshift space mocks (blue/red ellipses, respectively). The dotted ellipses show the constraints using the 1PT model of the velocity profiles, while the solid ones correspond to Eq.\ref{Eqvp2PT}. The semi-empirical correction of the velocity profiles is essential to recover the fiducial cosmologies.} 
\label{FigellRP}
\end{figure}

\medskip
Finally, we test what would be the effect of using Eq.\ref{Eq2PTok} and letting $\sigma_v$ as a free parameter in the MCMC, just like we do in the linear approximation. In Fig.\ref{FigellRP}, we can see the resulting constraints on $(f\sigma_8, \sigma_v)$ for the $\Lambda$CDM model (red ellipses), while the blue ellipses correspond to the RPCDM model. The dotted/solid ellipses correspond to the linear approximation (1PT)/our proposed model (Eq.\ref{Eq2PTok}) for the radial velocity respectively. The red/blue vertical dotted lines show the fiducial values of $f\sigma_8$ for $\Lambda$CDM/RPCDM models, while the horizontal lines correspond to the measured value of $\left<\sigma_v(R)\right>$ and $\sigma_{\rm eff}=0.9\left<\sigma_v(R)\right>$ over the fitting ranges $[0,75]\rm Mpc.h{-1}$.

\medskip
This illustrates how correcting the velocity profile with Eq.\ref{Eq2PTok} is the crucial ingredient to avoid systematic errors in the derivation of the growth rate. Without this correction, the inferred growth rate from the $\Lambda$CDM model has a systematic error that can be misinterpreted as the RPCDM cosmology. The inferred velocity dispersion is on the other hand in good agreement with the mean value we measured in the simulation, in all cases.

\section{Conclusion}\label{conclu}
In this work we investigated the limitation of the Gaussian streaming model applied to RSD around voids. We found that the simple linear theory prediction \cite{Peebles}, that has been previously used to probe the growth rate in galaxy surveys \citep{HamausSDSS,AB_voids2016,AH_Viper2016}, can not accurately reproduce the measured velocity profiles around voids. This can lead to systematic offsets in the derivation of the growth rate, as observed in \citep{AH_Viper2016,HamausSDSS2,IA2016}. 

\medskip
In this paper, we propose a semi-empirical correction (Eq.\ref{Eq2PTok}) to the velocity profiles, as a function of the underlying density contrast. This correction, combined with a calibrated velocity dispersion (not described as a constant), allows us to derive the growth rate without any systematic errors in our void samples. 

\medskip
 We also tested the characteristics of the velocities of dark matter particle distributions around voids, for a quintessence type of dark energy. Interestingly, we found that the critical imprints of quintessence, compared to the $\Lambda$CDM model, are captured by the velocity dispersion. The velocity dispersion can be seen as proportional to the mass assembly of matter, which is reduced for the RPCDM model due to a less decelerated expansion during the matter dominated era \cite{Alimi2010}. This warrants the treatment of $\sigma_v(R)$ as a physical parameter, rather than a nuisance parameter. However, we find that even by letting $\sigma_v$ as a free parameter, our proposed semi-empirical correction can successfully recover the fiducial cosmology, in contrast to the linear approximation, which induces a systematic offset in the growth rate derivation. 

\medskip
Finally, this work can be extended to different approaches. First, it would be interesting to see if our semi-empirical correction of the radial velocity prediction can correct the systematic offset in the growth rare measurement found in \cite{HamausSDSS2}, at low redshift. Second, in order to study more precisely the constraints on the growth rate using RSD around voids, it becomes essential to decompose the void-galaxy correlation function into multipoles (e.g. \cite{Cai2015}, \cite{HamausSDSS2}), and use the covariance matrix when performing the MCMC analysis. We hope to address this issue in a future work.

\medskip
\section*{Acknowledgements}
I would like to thank Prof. C. Blake for useful discussions, proofreading this manuscript as well as providing great support and inspiration. I also want to thank Prof. Y. Rasera for facilitating the access to the DEUS SIMULATIONS. This work was conducted by the Australian Research Council Centre of Excellence for All-sky Astrophysics (CAASTRO), through project number CE110001020. We also acknowledge support from the DIM ACAV of the Region Ile-de-France.

\bibliography{biblio_voids.bib}

\end{document}